\documentclass[aps,pre]{revtex4}
\usepackage[a4paper,left=25mm,top=20mm,right=30mm,bottom=15mm]{geometry}
\usepackage{amsmath}
\begin{document}
\title{A Strongly Coupled Open System with Non-linear Bath: \\ Fluctuation-Dissipation and Langevin Dynamics}
\author{Chitrak Bhadra}
\affiliation{Department of Physics, Jadavpur University, Kolkata, India}

\begin{abstract}
\noindent Study of Langevin dynamics and the fluctuation-dissipation relation (FDR) for a generic probe system (represented by a mass $M$), bilinearly coupled to a bath of harmonic oscillators, has been a standard paradigm for a microscopic theory of stochastic processes for several decades. The question that we probe in this paper is, how far the structure of the classical FDR is robust, when one replaces the harmonic bath by an anharmonic one in the limit of strong system-bath coupling? Such a picture carries the signature of the probe system in the zeroth order through a nonlocal time kernel. We observe that the two-time noise correlations hold out a rich structure from which the usual FDR emerges only in the leading order of perturbation. Beyond this order, multiple time scales and nontrivial dependence on the temperature starts manifesting. These new aspects conspire to break the time-translational invariance of the noise-correlations. Several other interesting features show up and we discuss them methodically through rigorous calculations order-by-order in perturbation. This formalistic derivation along with a specific example of non-linearity can be easily applied to a huge range of processes and statistical observables that fall under the purview of system-reservoir theory.
\end{abstract}

\maketitle
\section{Introduction}

\noindent The study of dissipative systems, classical and quantum, in the presence of random forces (noise) has been in focus for more than a century \cite{weiss}. At the classical level, the mathematical paradigm is based on the Langevin equation, a stochastic differential equation formulation. It describes the motion of a system under an external potential with two important ingredients: dissipation of energy over time and random noise, both arising due to interaction with an environment (we will call it a bath throughout the paper). In the limit of Gaussian white noise (Markovian limit) and an initial equilibrium thermal distribution of the bath modes, the Langevin equation describes Brownian Motion in the continuum and the diffusion co-efficient emerges as a natural parametrisation of the motion.On the other hand, Generalised Langevin Equations (GLE) have been investigated over the decades which explain effects of nonlinear damping and non-Markovian, non-Gaussian noise \cite{mori,zwa1,zwa2}. Also, a probabilistic description of the system dynamics is governed by the Fokker Planck equation that emerges directly from the GLEs and is an essential tool in calculating avergaed statistical observables \cite{risken}. The phenomenology explained by Langevin equations cover a vast range of stochastic dynamical phenomena in nature and dictate experiments at small length scales \cite{hang1,hang2,hang3,dhar}.\\

\noindent On a second look, a GLE describes the coarse-grained effect of the bath on the system itself. Therefore, a system-bath Hamiltonian approach was developed by Zwanzig to account for a microsopic description of classical GLEs \cite{weiss,mori,zwa1,zwa2}. Intuitively, the picture is simple: since dissipation and noise arise out of the interaction between system and the environment in which its dynamics occurs, an explicit partioning of the Hamiltonian is done. This contains the system, bath and interaction Hamiltonian systematically and subsequent equation of motion under a probabilistic interpretation of initial conditions for the bath lead to a coarse-grained dynamical equation representing a GLE.Most importantly, an explicit derivation of Fluctuation-Dissipation Relations (FDR) \cite{kubo} which relate the damping kernel with the noise correlations emerge naturally \cite{weiss,mori,zwa1,zwa2,cortes}. The spirit of system-bath partitioning can also be found in the seminal work of Lebowitz {\it et al} in understanding classical non-equilibrium Stationary States involving multiple thermal reservoirs \cite{leb}. At the quantum level, Feynman-Vernon used this system-bath viewpoint with great clarity to quantise dissipative systems \cite{feyn}. Caldeira-Leggett took the method a step forward in describing quantum Brownian motion (QBM) and derived a master equation \cite{legg,grab} for the evolution of the density-matrix of the system which have major implications in the theory of decoherence and other dissipative processes \cite{weiss,zurek,blen}.\\

\noindent Our present work addresses two subtle questions in system-bath formalism at the classical level: the presence of an intrinsic nonlinearity (represented by parameter $\epsilon$) in the bath modes (a thermal bath and uncoupled initial condition is assumed) and a linear but arbitrarily strong system-bath coupling ($\lambda$). Earlier,\textit{nonlinear system-bath coupling} has been investigated where state-dependent damping and coloured noise emerge naturally maintaining the robustness of FDR \cite{weiss}. At the quantum level, a systematic perturbative approach has been introduced to derive the quantum Master equation for such processes leading to significant implications in the theory of decoherence \cite{hu}. Also, \textit{nonlinear or periodic external potentials} are a playing field for many realistic physical processes like Brownian motors, barrier crossing dynamics, to name just a few \cite{hang3,ray1,ray2}. However, the thermal baths in all these models are always considered to be made up of uncoupled or coupled harmonic oscillator modes that indeed serve as a very good approximation for the forces exerted by a bath.\\

\noindent Recently, the author showed an added feature: the bath modes were considered nonlinear over and above their pure quadratic nature \cite{bhadra}. Firstly, mathematically this needed a perturbative approach to evaluate the GLE and the FDR obeyed by the noise correlations because the essential quadratic nature of the bath is lost. At the zeroth order, the harmonic term serves as the solution over which the perturbative expansion is done. This makes the noise kernel manifestly nonlinear in bath coordinates and the FDR was shown to pick up correction terms that intrinsically measure the excess correlations due to the nonlinearity. FDRs for double well baths and quartic baths have been explicitly calculated and an interpretation of the GLEs in terms of effective temperature dependent damping was elucidated. Such effective Langevin Equations and non-equilibrium FDRs have been reported in various studies recently \cite{krug,maes,marco,sarr,shi}. Nevertheless, we point to the fact that a crucial ingredient in working with nonlinear baths was the \textit{weak system-bath regime allowing a convenient perturbative formalism}. To the first order investigated in the coupling strength and parameter for nonlinearity, it was found \textit{that the simple velocity dependent damping term survives}; i.e., \textit{the all the nonlinearity in GLE shows up only in the noise-kernel}. It seems the system coupled to the quadratic part of the Hamiltonian robustly maintains the FDR. The nonlinearities (of uncoupled modes), representing a form of self-interaction of the bath modes, impose themselves as small periodic non-equilibrium fluctuations over the usual FDR.\\

\noindent In this work, we remove this special condition: i.e., the system-bath coupling is left arbitrarily large. Many recent attempts have been made to understand the fundamental implications of system-bath strong coupling in stochastic thermodynamics, non-equlibrium fluctuation relations, etc \cite{seif,jarz,stras,hu2,camp}. Here, the coupling in its functional form is assumed bilinear in system-bath coordinates to keep calculations tractable. The zeroth order solution for the bath modes are now \textit{sensitive to the state of the probe system through an usual nonlocal kernel}. The perturbative expansion is only restricted to the small nonlinearity in the bath modes. To illustrate the intricate interplay of the coupling and nonlinearity parameters, we initially evaluate the form of the GLE for a single bath mode (a cubic nonlinearity, the simplest case) and then smoothly transpose the formalism to a realistic many-particle bath.The first indicator of strong coupling shows up in extra pieces of nonlinear velocity and position dependent dissipation functions in the GLE. Naturally, it will be easy to see that the weak coupling scheme suppressed such terms.The noise has three parts: linear, nonlinear and nonlocal kernels. The FDR of the second kind is next derived via usual averaging methods under suitable probability distribution for the bath modes. As a further extension, we show multiple time-scales emerge in the damping and noise kernels which are a signature of the strong interaction of the system with the nonlinear part of the bath Hamiltonian.\\

\noindent At this point it will be useful to briefly highlight the key results of this paper. \textit{Firstly}, the GLE, which is the central equation of the entire discussion that follows, has been derived in Eq.(\ref{eq30}). With $M$ denoting the mass of the system, the GLE, in terms of its coordinate $X(t)$, looks like

\begin{equation}
M\ddot{X} + \lambda^{3} \epsilon \int_{0}^{t} dt' \Sigma^{nl}(t-t') + \lambda^{2} \int_{0}^{t} dt'\gamma^{l}(t-t')\dot{X}(t') + V'(X) = \Gamma _{\lambda}(t) +\Gamma _{\lambda \epsilon}(t) + \Gamma _{\lambda^{2} \epsilon}(t).
\label{eq00}
\end{equation}

\noindent Here $\epsilon$ and $\lambda$ are parameters attached to the nonlinearity of the bath modes and the coupling of the system (probe) to the bath modes respectively. While the entire analysis is based on a perturbative expansion in the small nonlinearity-parameter $\epsilon$, the emphasis of the formalism presented in this paper centres around the coupling parameter $\lambda$ \textit{not} being small, and hence the coupling term remains present right from the very beginning of the analysis, i.e., in the zeroth order equation in the bath modes [see Eqs.(\ref{eq27}) and (\ref{eq28}) below]. The other terms present in the left hand side of the above equation are: (i) $\gamma^l (t - t')$, which denotes the damping term that constitutes the usual memory kernel along with a linear term in velocity [see Eq.(\ref{eq32}) below], (ii) $\Sigma^{nl} (t - t')$, which is a memory kernel dependent on nonlinear powers of the system's position and velocity [see Eqs.(\ref{eq33}), (\ref{eq31}) and (\ref{eq21}) below] and (iii) gradient of some potential $V(X)$. On the right hand side of the above equation, various noise terms appear. Even within the tenets of a first order perturbative calculation in the perturbation parameter $\epsilon$, appearance of various powers of the non-perturbative coupling parameter $\lambda$ becomes inevitable. Their mathematical structures have been written down in the Appendix [see Eqs.(\ref{eq57a})-(\ref{eq57c})] and in a slightly simplified form in Eqs.(\ref{eq19a})-(\ref{eq19c}). Being designated as noise terms, they consist of complicated combinations of initial values of the dynamical variables describing the bath modes. \textit{Secondly}, the generalized FDR, that has been schematically written in Eq.(\ref{eq43}) as a sum of several two-time noise correlations involving the second, third and fourth powers of the non-perturbative parameter $\lambda$, looks like

\begin{eqnarray}
\langle\Gamma(t)\Gamma(t') \rangle &=& \langle \Gamma _{\lambda}(t)\Gamma _{\lambda}(t')\rangle + \langle \Gamma _{\lambda \epsilon}(t)\Gamma _{\lambda \epsilon}(t')\rangle_{0} + \langle \Gamma _{\lambda^{2} \epsilon}(t)\Gamma _{\lambda^{2} \epsilon}(t')\rangle_{0}\nonumber\\
&&+ \langle \Gamma _{\lambda}(t)\Gamma _{\lambda \epsilon}(t')\rangle + \langle \Gamma _{\lambda}(t)\Gamma _{\lambda ^{2} \epsilon}(t')\rangle + \langle \Gamma _{\lambda \epsilon}(t)\Gamma _{\lambda^{2} \epsilon}(t')\rangle_{0} + (t \leftrightarrow t').
\label{eq01}
\end{eqnarray}

\noindent The various terms appearing on the right hand side of this equation involve different combinations of powers of $\lambda$ and $\epsilon$ [see Eqs.(\ref{eq44})-(\ref{eq49}) below]. For example, the first term on the right hand side of Eq.(\ref{eq01}) is of the order of ($\lambda^2$), the second ($\lambda^2.\epsilon^2$), the third ($\lambda^4.\epsilon^2$), the fourth ($\lambda^2.\epsilon^2$), the fifth ($\lambda^3.\epsilon$) and the sixth evaluates to zero. The first term, which goes as the quadratic power of the coupling constant and carries no footprint of the bath-nonlinearity, corresponds to the usual FDR obtained in the standard Zwanzig formalism with harmonic baths. The other terms bear the signatures of the bath-nonlinearity to various extents. Even within first order in $\epsilon$, several time-kernels appear and dependence of the two-time noise correlation at this level is characterized by its dependence on system variables as well [see Eqs.(\ref{eq45}) and (\ref{eq34}) below]. The other terms, which are of the order of $\epsilon^2$, involve complicated interplay of higher harmonics of the bath frequencies and also carry explicit signatures of breaking of time-translational invariance.\\

\noindent The paper has been organised as follows: in Section II the system-bath model is introduced and perturbative techniques employed to evaluate the classical equation of motion for a single bath particle. A schematic GLE shows the importance of each term order by order in two relevant parameters: the system-bath coupling and small nonlinearity. In Section III we generalize to a realistic bath of large number of bath modes and multiple time-scales and nonlinearities are analysed in the context of the resulting GLE. Here, a subsection leads into the calculation of noise-correlations. The FDR of the second kind is shown to be robust in the lowest orders of perturbation in $\epsilon$ and a temperature dependent effective damping kernel is motivated due to extra higher order correlations carrying signature of nonlinearity. Section IV puts forward the strongest case for studying such nonlinear baths with arbitrary coupling strength and how they relate to other extensions of Caldeira-Leggett model.\\

\section{System coupled to a nonlinear oscillator: a cubic perturbation}

\noindent In this section we introduce the dynamics of a generic system coupled to one oscillator mode of the bath in the same spirit as the Caldeira-Leggett model . Since finally we will resort to a realistic bath mode of a distribution of nonlinear but uncoupled oscillators, the following calculations will reveal the possibilities we have already laid down.

\noindent \subsection{System interaction with single bath mode}

\noindent The total Hamiltonian is naturally defined as

\begin{equation}
H_{total} = H_{S}+H_{B}+H_{SB}+H_{CT}
\label{eq1}
\end{equation}

\noindent The first two terms on the right hand side respectively represent the system and bath Hamiltonians, the third term represents the interaction potential between the system and the bath and finally, the last term is a counter-term that cancels spurious potential renormalsiation in the classical scenario that will be implemented naturally later in the paper. The Hamiltonian of the anharmonic oscillator, with coordinate $q$, is parametrised by the natural frequency $\omega$ and perturbation parameter $\epsilon$.

\begin{eqnarray}
H_{S} &=& \frac{P^{2}}{2M}+V(X)
\label{eq2} \\
H_{B} &=& \frac{p^{2}}{2m} + \frac{1}{2}m\omega^{2}q^{2} + \epsilon.\frac{1}{3}q^{3}.
\label{eq3}
\end{eqnarray}

\noindent The interaction Hamiltonian is defined by the coupling strength $\lambda$ that is assumed to be an arbitrarily valued parameter. The coupling is bilinear in the respective position coordinates and can be written as

\begin{equation}
H_{SB} = -\lambda q(t)X(t).
\label{eq4}
\end{equation}

\noindent We shall later see that the sign of the parameter $\lambda$ plays a significant role in the open system dynamics under the influence of an anharmonic bath.\\

\noindent  At this point, it is crucial to highlight that due to the heaviness of the mass of the system (denoted $M$) in comparison to the mass of the reservoir oscillator (denoted $m$), the former does not allow the latter to roll down to negative infinity along the negative arm of the cubic potential. This is an important role also played by the coupling between the system and the bath oscillators, subject to the fact that in Eq.(\ref{eq3}) the nonlinearity in the potential of the bath oscillator is sufficiently weak compared to the harmonic part.


\subsection{Solution for bath mode}   

\noindent The equations of motion for the system and bath mode which are coupled through the parameter $\lambda$ are:

\begin{eqnarray}
M\ddot{X} &=& -V^{'}(X) + \lambda q(t)
\label{eq5} \\
m\ddot{q} &=& -m \omega ^{2} q + \lambda X(t) - \epsilon q^{2}.
\label{eq6}
\end{eqnarray}

\noindent The usual strategy now is to solve for the bath mode and replace it in the original system equation of motion thereby integrating out the bath degree of freedom. For arbitrarily strong coupling, the perturbation theory for the nonlinear bath has to be done systematically. A formal solution for $q(t)$ reads:

\begin{eqnarray}
q(t) &=& q^{(0)} + \epsilon q^{(1)} + O(\epsilon ^{2})
\label{eq7} \\
\omega &=& \omega^{(0)} + \epsilon \omega ^{(1)} + O(\epsilon ^{2}).
\label{eq8}
\end{eqnarray}

\noindent Standard perturbation theory dictates that the natural frequency of the harmonic mode of the bath should also be expanded as it receives correction from higher orders \cite{bhadra}. To zeroth order, the equation reveals a harmonic oscillator under the influence of a forcing term which \textit{by virtue of the bilinear coupling is equal to the system co-ordinate}.

\begin{eqnarray}
m\ddot{q}^{(0)}(t) &=& - m{[\omega^{(0)}]}^2 q^{(0)} + \lambda X(t)
\label{eq9} \\
q^{(0)}(t) &=& q^{(0)} (0) \cos (\omega^{(0)}t) + \frac{p^{(0)} (0)}{m \omega^{(0)}} \sin (\omega^{(0)}t) + \frac{\lambda}{m \omega^{(0)}} \int_{0}^{t} dt{'}\sin (\omega^{(0)}(t-t{'}))X(t{'}).
\label{eq10}
\end{eqnarray}

\noindent This zeroth order evolution of the bath mode bears significant differences from that of weak coupling to nonlinear baths. Firstly, the coupling $\lambda$ sits explicitly in the solution along with the nonlocal system induced kernel; the position of the probe affects the bath mode non-perturbatively. This single equation provides the starting point from which the whole subsequent analysis that will follow. Any small nonlinearity in the bath is expanded over and above this inhomogeneous solution. The equation in first-order of perturbation reads:

\begin{equation}
m\ddot{q}^{(1)} (t) = - m{[\omega^{(0)}]}^2 q^{(1)}(t)  -2 m\omega^{(0)} \omega ^{(1)}q^{(0)}(t)- \epsilon {[q^{(0)}(t)]}^2,
\label{eq11}
\end{equation}

\noindent whose solution can, as usual, be partitioned into a complementary function and a particular solution.

\begin{equation}
q^{(1)} = q_{c}^{(1)} + q_{p}^{(1)}.
\label{eq12}
\end{equation}

\noindent Again, the superscript for the perturbed co-ordinate and frequency are maintained explicitly. Solving in a similar fashion as above, the entire solution can be written formally as:

\begin{eqnarray}
q^{(1)}(t)&=& \Big(q^{(1)} (0) \cos (\omega^{(0)}t) + \frac{p^{(1)} (0)}{m \omega^{(0)}} \sin (\omega^{(0)}t)\Big) - \Big(2 \omega ^{(1)} \int_{0}^{t} dt' q^{(0)}(t')\sin (\omega^{(0)}(t-t'))
\nonumber \\
&&- \frac{\epsilon}{m \omega^{0}} \int_{0}^{t} dt' {[q^{(0)}(t')]}^2 \sin(\omega^{(0)}(t-t')\Big).
\label{eq13}
\end{eqnarray}

\noindent The nonlinearity in the last term of Eq.(\ref{eq11}) now will be expanded and a nonlocal kernel $F(t)$ will be defined as it will recur throughout the calculation:

\begin{eqnarray}
{[q^{(0)} (t)]}^2 &=& A_{0} + A_{1} \cos (2\omega^{(0)}t) + A_{2} \sin (2\omega^{(0)}t)
\nonumber \\
&&+ \frac{2\lambda}{m\omega^{(0)}} F(t) \Big(q^{(0)} (0) \cos (\omega^{(0)}t) +  \frac{p^{(0)} (0)}{m \omega^{(0)}} \sin (\omega^{(0)}t)\Big) +\frac{\lambda^{2}}{m^{2}{{[\omega^{(0)}]}^2}} F^{2}(t)
\label{eq13a}
\end{eqnarray}

\noindent where,

\begin{equation}
F(t) = \int_{0}^{t} dt' \sin(\omega^{(0)}(t-t'))X(t').
\label{eq14}
\end{equation}

\noindent The factors $A_{0}$, $A_{1}$ and $A_{2}$ carry the nonlinear terms in the bath co-ordinate, $q^{(0)}$ and $p^{(0)}$ [see Appendix, Eqs.(\ref{eq56a})-(\ref{eq56c})]. This is a feature that occurs in any sort of nonlinearity in the bath and ultimately renders the noise to be nonlinear and multiplicative.\\

\noindent At this point it is worthwhile to investigate the meaning of the terms on the right hand side of Eq.(\ref{eq13}). As is well known in classical perturbation theory, expansion of these form inherently give rise to \textit{secular terms} which make the long-time solution divergent. To avoid such a possibility, the already expanded natural frequency of the harmonic oscillator comes into play. The trick is to identify the \textit{co-efficients of the secular pieces} and equate them to zero forming a constraint equation. Generally, these constraint equations carry the corrections, order by order, to the $\omega$ and also amplitude of the harmonic solution. Here, we pick up the third term on the right hand side of Eq.(\ref{eq13}), and sieve out the integrand that will produce long-time divergence.

\begin{eqnarray}
q^{(0)}(t')\sin (\omega^{(0)}(t-t{'})) &=& q^{(0)} (0) \cos (\omega^{(0)}t{'})\sin (\omega^{(0)}(t-t')) + \frac{p^{(0)} (0)}{m \omega^{(0)}} \sin (\omega^{(0)}t') \sin(\omega^{(0)}(t-t'))
\nonumber\\
&&+ \frac{\lambda}{m\omega^{(0)}}\sin (\omega^{(0)}(t-t')) \int_{0}^{t} ds\sin (\omega^{(0)}(t-s))X(s).
\label{eq15}
\end{eqnarray}

\noindent It is sufficient to take up the first term,

\begin{equation}
\int_{0}^{t} dt' q^{(0)} (0) \cos (\omega^{(0)}t')\sin (\omega^{(0)}(t-t')) = \frac{1}{2} \int_{0}^{t}  dt' q^{(0)}(0) \big(\sin (\omega^{(0)}(t)) - \sin (\omega^{(0)}(2t'-t))\big)
\nonumber
\end{equation}

\noindent and clearly, the first integral diverges. The second integrand in Eq.(\ref{eq15}) also suffers from the same fate. This leads immediately to the well known result for cubic nonlinearity,

\begin{equation}
\omega^{(1)} = 0
\label{eq16}
\end{equation}

\noindent i.e., the first order correction to the natural frequency of the oscillator is zero. To check for corrections, calculations to the second order have to be done. For completeness, we mention that the third term on the right hand side of Eq.(\ref{eq13}) involves either higher harmonics or the nonlocal system kernel thereby giving rise to no secular divergence.\\

\noindent
Now, we are ready to write down the complete solution for the first-order perturbation solution for the anharmonic bath mode coupled to the system. On the basis of the above analysis, we provide below a few simplified notations for convenience in doing the calculations that follow.

\begin{eqnarray}
q(t)&=& q^{(0)}(t) + \epsilon \big(q_{c}^{(1)} + q_{p} ^{(1)}\big)\nonumber\\
q(0)&=& q^{(0)}(0) + \epsilon q^{(1)}(0)\nonumber\\
p(0)&=& p^{(0)}(0) + \epsilon p^{(1)}(0)\nonumber\\
\omega&=&\omega^{(0)} \nonumber
\end{eqnarray}

\noindent The subtle point to be realised is that the replacement of the superscript $^{(0)}$ picks up an error in the subsequent calculation but only of the order $\epsilon ^{2}$ and hence is neglected. Using Eq.(\ref{eq10}) and solution of Eq.(\ref{eq13}) (with condition Eq.(\ref{eq16}) implemented), it follows:

\begin{eqnarray}
q(t) &=& q(0) \cos \omega t + \frac{p(0)}{m \omega} \sin \omega t + \frac{\lambda}{m \omega} \int_{0}^{t} dt'\sin (\omega(t-t'))X(t') \nonumber\\
&&+ \frac{\epsilon}{m\omega} \Big[B_{0}\Big(\cos\omega t - 1 \Big) + B_{1}\Big(\sin3\omega t +\frac{2}{3}\sin2\omega t -\sin\omega t\Big) \nonumber \\
&&+ B_{2}\Big(\cos3\omega t +\frac{2}{3}\cos2\omega t -\cos\omega t\Big)\Big] \nonumber\\
&& -\frac{2\lambda \epsilon}{m^{2}\omega^{2}}\Big[\int_{0}^{t} dt'\sin (\omega (t-t')) F(t')\Big( q(0) \cos \omega t' + \frac{p(0)}{m \omega}\sin \omega t' \Big) \Big] \nonumber \\
&&- \frac{\lambda^{2} \epsilon}{m^{3}{\omega}^{3}}\int_{0}^{t} dt' \sin (\omega (t-t'))F^{2}(t').
\label{eq17}
\end{eqnarray}

\noindent Here again, the coefficients $B_{0}$, $B_{1}$ and $B_{2}$ carry the signatures of nonlinearity in the initial values of bath coordinates [see Appendix, Eqs.(\ref{eq56d})-(\ref{eq56f})]. Noticeably, the terms of the order $\lambda \epsilon$
and $\lambda ^{2} \epsilon$ are the result of a \textit{non-perturbative treatment of the system-bath coupling}.


\subsection{A schematic Generalised Langevin Equation}

\noindent The solution given by Eq.(\ref{eq17}) when substituted back into the equation of motion for the system, i.e., Eq.(\ref{eq5}), we get the following equation that contains \textit{coarse-grained information about the bath nonlinearity} as well as the \textit{arbitrary system-bath coupling}:

\begin{eqnarray}
M\ddot{X} + \frac{\lambda^{3} \epsilon}{m^{3}\omega^{3}}\int_{0}^{t} dt' \sin (\omega (t-t'))F^{2}(t') - \frac{\lambda^{2}}{m \omega} \int_{0}^{t} dt'\sin (\omega(t-t'))X(t') +V'(X) &=&  \Pi_{\lambda} +\Pi_{\lambda \epsilon} + \Pi_{\lambda^{2} \epsilon} \nonumber\\
\label{eq18}
\end{eqnarray}

\noindent where,

\begin{eqnarray}
\Pi_{\lambda}(t) &=& \lambda\Big(q(0) \cos \omega t + \frac{p(0)}{m \omega} \sin \omega t\Big) 
\label{eq19a} \\
\Pi _{\lambda \epsilon}(t) &=& \frac{\lambda\epsilon}{m\omega} \Big[ B_{0}\Big(\cos\omega t - 1 \Big) + B_{1}\Big(\sin3\omega t +\frac{2}{3}\sin2\omega t -\sin\omega t\Big) \nonumber \\
&&+ B_{2}\Big(\cos3\omega t +\frac{2}{3}\cos2\omega t -\cos\omega t \Big)\Big] 
\label{eq19b} \\
\Pi _{\lambda^{2} \epsilon}(t) &=& -\frac{2\lambda^{2} \epsilon}{m^{2}\omega^{2}}\Big[\int_{0}^{t} dt'\sin (\omega (t-t')) F(t')\Big( q(0) \cos \omega t' + \frac{p(0)}{m \omega}\sin \omega t' \Big)\Big].
\label{eq19c}
\end{eqnarray}

\noindent A more familiar form of the left hand side of Eq.(\ref{eq18}) clearly elucidating the velocity dependent dissipation terms can be written down by doing by-parts integration and noting that $F(0)=0$ and $F(t)=0$:

\begin{eqnarray}
&& M\ddot{X} + \frac{2\lambda^{3} \epsilon}{m^{3}\omega^{4}}\int_{0}^{t} dt' \cos (\omega (t-t'))F(t')\dot{F}(t') + \frac{\lambda^{2}}{m \omega^{2}} \int_{0}^{t} dt'\cos (\omega(t-t'))\dot{X}(t') +V'(X) \nonumber \\
&=&  \Pi_{\lambda} + \Pi_{\lambda \epsilon} + \Pi_{\lambda^{2} \epsilon}.
\label{eq20}
\end{eqnarray}

\noindent The usual effect of a counter-term \cite{weiss} along with the initial value $X{(t{'}=0)}$ has been assumed.\\

\noindent Though the equation has been evaluated for only a single bath mode, several points warrant emphasis here. Firstly, it gives a schematic window to what the actual GLE will look like when a realistic many-particle bath is included. On the left hand side of Eq.(\ref{eq20}), a significant effect of a $\lambda ^{3} \epsilon$ term arising due to arbitrary system-bath coupling is observed: generation of a damping kernel which is nonlinear in system position and velocity and having a complicated nonlocal time dependence attached to it. If compared to the GLE of reference \cite{bhadra}, this term was naturally suppressed due to small $\lambda$ approximation. The third term on the right hand side of Eq.(\ref{eq20}) is the usual velocity dependent kernel that arises in any form of Caldeira-Leggett model or its extensions. Also, the right hand side has been categorised order by order in Eqs.(22)-(24), anticipating the interpretation of noise. We see the emergence of a state-dependent nonlocal kernel $\Pi _{\lambda^{2} \epsilon}(t)$ (coloured noise) due to the system-bath coupling being strong. A small $\lambda$ necessarily hides this term.\\


\noindent We further explore the kernel $F(t')\dot{F}(t')$ which sheds light on this new nonlinear damping term. Also, we neglect any term arising due to $X(t=0)$ that is taken to be at the origin without any loss of generality. Integrating Eq.(\ref{eq14}) by parts,

\begin{eqnarray}
F(t') &=& \int_{0}^{t'} ds X(s)\sin (\omega(t'-s)) \nonumber\\
&=& \frac{X(t')}{\omega} + \frac{1}{\omega} \int_{0}^{t'} ds \dot{X}(s)\cos (\omega(t'-s)) \nonumber
\end{eqnarray}

\noindent and after some algebra, we arrive at

\begin{eqnarray}
F(t')\dot{F}(t') &=& \frac{2}{\omega^{2}} X(t') \dot{X}(t') + \frac{2}{\omega^{2}}\dot{X}(t')\int_{0}^{t'} ds \dot{X}(s)\cos (\omega(t'-s)) - \frac{1}{\omega}X(t')\int_{0}^{t'} ds \dot{X}(s)\sin (\omega(t'-s)) \nonumber \\
&&- \frac{1}{\omega}\Big(\int_{0}^{t'} ds \dot{X}(s)\cos (\omega(t'-s)) \Big) \Big( \int_{0}^{t'} du \dot{X}(u)\cos (\omega(t' - u)) \Big)
\label{eq21}
\end{eqnarray}

\noindent This equation reveals the inherent nonlinear character of the new dissipative term. The double time kernel manifestly will break the generic $K(t-t{'})$ form of the dissipation factor in the GLE and such a situation will again arise when the FDR is calculated shortly. Finally, we resort to a well known limit in system-bath dynamics, \textit{the slow probe limit}. Due to the heavy mass of the system particle relative to the bath mode, it is assumed that its velocity $\dot{X}(t)$ changes very slowly during time-intervals over which the damping kernel is calculated. Moreover, $\frac{1}{\omega}$ sets the fastest time-scale for the bath dynamics and the system-bath motion as a whole.

\begin{equation}
\frac{1}{\omega}<<\frac{X(t)}{\dot{X}(t)}.
\label{eq22}
\end{equation}

\noindent Therefore the velocity term can be shifted out of the integrals thus leading to a much more recognizable Langevin equation for short time scales. In the slow-probe limit the schematic GLE becomes:

\begin{eqnarray}
M\ddot{X} + \frac{\lambda^{3} \epsilon}{m^{3}{\omega}^{3}}\int_{0}^{t} dt' \Omega(t') + \frac{\lambda^{2}}{m \omega^{2}} \int_{0}^{t} dt' \cos (\omega(t-t'))\dot{X}(t') +V'(X) &=&  \Pi_{\lambda} +\Pi_{\lambda \epsilon} + \Pi_{\lambda^{2} \epsilon}
\label{eq23}
\end{eqnarray}

where,

\begin{equation}
\Omega(t') = \frac{1}{\omega^{3}}(1- K(t,t'))\dot{X^{2}}(t').
\label{eq24}
\end{equation}

\noindent The kernel with variables $t$ and $t'$ are expanded in the Appendix for the many-particle case discussed in the next Section [see Appendix, Eqs.(\ref{eq57a})-(\ref{eq57c})]. However, this form already shows clearly the kind of dissipation expected in a nonlinear bath model when the coupling to the system is kept arbitrarily large. The position and velocity of the system intrinsically gets linked with the noise variables and \textit{produce quadratic velocity-dependent terms}, something which we will show as impossible with quadratic baths and generic couplings in Section III.

\section{The full system-bath model and noise-correlations}

\noindent To describe the dynamics of a dissipative system and extract the statistical mechanical consequences of the system-bath interaction, we need to transpose the whole formalism above to a many-particle environment. A discrete set of nonlinear oscillators (later promoted to a continuous distribution) define the effect of the bath. The initial conditions of these bath particles are assumed to be distributed according to a classical canonical ensemble defined via temperature $T$, i.e., in equilibrium. As they are mutually uncoupled, every effect of the bath modes can be obtained by summing over the single mode calculations. Moreover, even if they were coupled through harmonic interactions, a similarity transform would lead to normal coordinates which would again represent virtual uncoupled modes. Under such a paradigm, we can finally reach a compact yet nonlinear and nonlocal form of the GLE of the system in question (Section A). As a next step, the average and two-time noise-correlations are rigorously calculated giving rise to a generalised form of FDR of the second kind (Section B). The central results are best summarised in Eqs.(\ref{eq30}), (\ref{eq42}) and (\ref{eq43})-(\ref{eq49}) below.

\subsection {Generalised nonlinear Langevin Equation}

\noindent The Hamiltonian for the system-bath dynamics become:

\begin{eqnarray}
H_{S} &=& \frac{P^{2}}{2M}+V(X)\nonumber\\
H_{B} &=& \sum_{\mu=1}^{N} \frac{p_{\mu}^{2}}{2m_{\mu}} + \sum_{\mu=1}^{N} \frac{1}{2}m_{\mu}\omega_{\mu}^{2} q_{\mu}^{2}+ \epsilon \sum_{\mu=1}^{N}\frac{q_{\mu}^{3}}{3}\nonumber\\
H_{SB}&=& -\lambda \sum_{\mu=1}^{N} c_{\mu}q_{\mu}X.
\label{eq25}
\end{eqnarray}

\noindent The equations of motion follow:

\begin{eqnarray}
M\ddot{X} &=& -V^{'}(X)  +\lambda \sum_{\mu=1}^{N} c_{\mu} q_{\mu}(t)
\label{eq26} \\
m\ddot q_{\mu}&=& -m \omega ^{2} q_{\mu} + \lambda c_{\mu}X(t) - \epsilon q_{\mu}^{2}.
\label{eq27}
\end{eqnarray}

\noindent The solution for the $\mu$-th mode of the bath is given by rewriting Eq.(\ref{eq17}) with proper subscripts:

\begin{eqnarray}
q_{\mu}(t) &=& q_{\mu}(0) \cos \omega_{\mu} t + \frac{p_{\mu}(0)}{m_{\mu} \omega_{\mu}} \sin \omega_{\mu} t + \lambda \frac{c_{\mu}}{m_{\mu} \omega_{\mu}} \int_{0}^{t} dt'\sin (\omega_{\mu}(t-t'))X(t')
\nonumber\\
&&+ \epsilon \frac{1}{m_{\mu}\omega_{\mu}} \Big[B_{\mu0}\Big(\cos\omega_{\mu} t - 1 \Big) + B_{\mu1}\Big(\sin3\omega_{\mu} t +\frac{2}{3}\sin2\omega_{\mu} t -\sin\omega_{\mu} t\Big)
\nonumber \\
&&+ B_{\mu2}\Big(\cos3\omega_{\mu} t +\frac{2}{3}\cos2\omega_{\mu} t -\cos\omega_{\mu} t\Big)\Big]
\nonumber\\
&&- \lambda \epsilon \frac{2c_{\mu}}{m_{\mu}^{2}\omega_{\mu}^{2}}\Big[\int_{0}^{t} dt' \sin (\omega_{\mu} (t-t')) F(t')\Big( q_{\mu}(0) \cos \omega_{\mu} t' + \frac{p_{\mu}(0)}{m \omega}\sin \omega_{\mu} t'\Big)\Big]
\nonumber \\
&& -\lambda^{2} \epsilon \frac{c_{\mu}^{2}}{m_{\mu}^{3}{\omega}_{\mu}^{3}}\int_{0}^{t} dt' \sin (\omega_{\mu} (t-t'))F^{2}(t').
\label{eq28}
\end{eqnarray}

\noindent Here,

\begin{equation}
B_{\mu i} \equiv B_{i}\rightarrow B_{\mu i}.
\label{eq29}
\end{equation}

\noindent Finally, the GLE can be constructed in a compact form analogous to Eq.(\ref{eq20}) as,

\begin{equation}
M\ddot{X} + \lambda^{3} \epsilon \int_{0}^{t} dt' \Sigma^{nl}(t-t') + \lambda^{2} \int_{0}^{t} dt'\gamma^{l}(t-t')\dot{X}(t') + V'(X) = \Gamma _{\lambda}(t) +\Gamma _{\lambda \epsilon}(t) + \Gamma _{\lambda^{2} \epsilon}(t)
\label{eq30}
\end{equation}

\noindent The right hand side of Eq.(\ref{eq30}) now defines noise (under a suitable probabilistic interpretation) in the proper sense as it is a summation over all the bath modes and mathematically equivalent to the form of the $\Pi$-terms defined in Eq.(\ref{eq19}) but properly summed over and subscripted with $\mu$. The generalised form of the integral given in Eq.(\ref{eq14}) will be called $G_{\mu}(t)$ instead of $F_{\mu}(t)$ in what follows [see Appendix, Eqs.(\ref{eq57a})-(\ref{eq57c})]. The redefinition of certain terms in going from single bath mode to multiple modes are summarised below:

\begin{eqnarray}
\lambda &\rightarrow& c_{\mu}\lambda
\nonumber\\
\Gamma_\alpha &\equiv& \Pi_\alpha\rightarrow \sum_{i=1}^{N}\Pi_{\mu \alpha}
\nonumber\\
G_{\mu}(t) &\equiv& F(t)\rightarrow F_{\mu}(t) 
= \int_{0}^{t} dt' X(t')\sin (\omega_{\mu}(t-t')).
\label{eq31}
\end{eqnarray}

\noindent In Eq.(\ref{eq30}), the usual linear-velocity dependent damping term leads to the standard kernel of the form

\begin{equation}
\gamma^{l}(t-t') = \sum_{\mu=1}^{N}\frac{c_{\mu}^{2}}{m_{\mu} \omega_{\mu}^{2}}\cos (\omega_{\mu}(t-t'))
\label{eq32}
\end{equation}

\noindent where the superscript ``l" denotes \textit{linear}. The nonlinear velocity-position dependent dissipative function which is intrinsically mixed with the bath modes due to strong coupling becomes:

\begin{equation}
\Sigma^{nl}(t-t') = 2\sum_{\mu=1}^{N}\frac{c_{\mu}^{3}}{m_{\mu}^{3}\omega_{\mu}^{4}}\cos (\omega_{\mu} (t-t'))G_{\mu}(t')\dot{G}_{\mu}(t').
\label{eq33}
\end{equation}

\noindent Here the superscript ``nl" denotes \textit{nonlinear}. For future use, we also define damping kernel mixing bath and system influences as, in analogy with Eq.(\ref{eq32}) as,

\begin{equation}
\gamma^{nl}_{\mu}(t-t') = \frac{2c_{\mu}^{3}}{m_{\mu}^{3}\omega_{\mu}^{4}}\cos (\omega_{\mu} (t-t'))G_{\mu}(t').
\label{eq34}
\end{equation}

\noindent The term defined in Eq.(\ref{eq34}) will appear in the formulation of the FDR later.\\

\noindent The above Eq.(\ref{eq30}) can be considered the central result of this paper. The effect of weakly perturbed nonlinear bath on a strongly coupled generic system has been thus laid down. The noise terms on the right hand side of Eq.(\ref{eq30}) exhibit the usual linear, weakly nonlinear and state-dependent character in three separate orders of the parameters determining the dynamics. On the left hand side the usual damping kernel, dependent linearly on velocity, is complemented with a nonlinear velocity as well as position dependent dissipation function. Strikingly, the fact that the system bath coupling is arbitrarily large has led to the intrinsic mixing of the dynamical quantities of the system and the bath. \textit{No longer can this term be separated out into a form where the effect of the bath and the velocity and position of the system are product separable.} For consistency check, we note that when the bath nonlinearity is absent, i.e., $\epsilon=0$, the GLE recovers the standard form with only a linear-velocity dissipation and linear noise. Moreover, in the limit $\lambda<1$, i.e., when the system-bath coupling is treated perturbatively, the higher order nonlinearities are suppressed and nonlinear noise terms arise. This leads to an effective form of the GLE with temperature-dependent damping, as elucidated in reference \cite{bhadra}.


\subsection{Generalised Fluctuation-Dissipation Relation}

\noindent In this section we evaluate the noise-correlations of the open system dynamics. These statistical averages provide a window into the coarse-grained effect of bath on the generic probe system. As mentioned above, for quadratic baths a robust FDR follows as a crucial step in equating the dissipative friction and thermal noise justifying strongly the microscopic picture that leads to a GLE. Such robustness is present in nonlinear system-bath coupling schemes too \cite{cortes}. Even for weak system-bath coupling and nonlinear baths, it has been shown \cite{bhadra} that at the lowest order of perturbation in $\lambda$, the structure of the FDR remains intact. Also, the FDR involves only natural time scales of the bath oscillators, $\frac{1}{\omega_{\mu}}$ and \textit{no higher harmonics of $\omega_{\mu}$}; along with this, the time translation invariant form of the kernel $K(t-t')$  is also retained.\\

\noindent As opposed to the aforementioned aspects in regard to the robustness in the structure of the FDR for closed systems, in the present work, we find that the \textit{averages differ significantly} due to the cubic nature of nonlinearity and also the strong coupling between the system  and the bath. Generalised FDR of the second kind relates the damping and the noise. Excess fluctuations violate the time translation invariance of usual FDR and, consequently, multiple time-scales emerge in the correlation functions. We recall from Eq.(\ref{eq30}), the noise has three separate terms [see Appendix, Eqs.(\ref{eq57a})-(\ref{eq57c})]:

\begin{equation}
\Gamma(t) = \Gamma _{\lambda}(t) +\Gamma _{\lambda \epsilon}(t) + \Gamma _{\lambda^{2} \epsilon}(t)\nonumber
\end{equation}

\noindent The first term carries the simple linear noise term due to quadratic bath. The second is due to the small cubic nonlinearity that can be treated perturbatively. The final term carries the signature of non-perturbative system-bath coupling. The initial bath modes described by the sets of bath variables $\{q_{\mu}\}$ and $\{p_{\mu}\}$ are assumed to be distributed according to a thermal distribution (a canonical ensemble) defined with respect to an absolute temperature $T$. We note that the bath Hamiltonian is nonlinear and so the partition function ($Z'$) and the probability distribution $P(\{q_{\mu}(0)\},\{p_{\mu}(0)\})$ of the initial values also carry the information of this nonlinearity. Thus an expansion around the quadratic part of the partition function $Z_{0}$ is natural [see Appendix, Eqs.(\ref{eq58})-(\ref{eq60})].

\begin{equation}
Z' = Z_{0}\Big(1 - \beta \epsilon \langle{H_{n}}\rangle_{0}\Big).
\label{eq35}
\end{equation}

\noindent From Eq.(\ref{eq25}), the quadratic and nonlinear Hamiltonians can be distinguished as,

\begin{eqnarray}
H_{0} &=& \sum_{\mu=1}^{N} \Big(\frac{p_{\mu}^{2}}{2m_{\mu}}+\frac{1}{2} m_{\mu}\omega_{\mu}^{2} q_{\mu}^{2}\Big)
\label{eq36} \\
H_{n} &=& \sum_{\mu=1}^{N} \frac{q_{\mu}^{3}}{3}
\label{eq37}
\end{eqnarray}

\noindent and with the definitions

\begin{eqnarray}
\langle{...}\rangle_{0} &\equiv& \frac{1}{Z_{0}}\int d\Omega \Big(...\Big) \exp \Big(-\beta H_{0}\Big)\nonumber\\
d\Omega &\equiv& \Big(\prod_{\mu =1}^{N}\prod_{\mu =1}^{N}dp_{\mu}(0)dq_{\mu}(0)\Big)\nonumber
\end{eqnarray}

\noindent the cubic nature of the perturbation leads to

\begin{eqnarray}
\langle{H_{n}}\rangle_{0}&=&0
\label{eq38} \\
Z{'}&=&Z_{0}.
\label{eq39}
\end{eqnarray}

\noindent Therefore, the canonical ensemble for the initial conditions of the cubic bath oscillators \textit{that will be used for calculating all the averages henceforth} reduces to:

\begin{equation}
P'(\{q_{\mu}(0)\},\{p_{\mu}(0)\}) = (1 - \beta\epsilon H_{n})P_{0}(\{q_{\mu}(0)\},\{p_{\mu}(0)\}) + O(\epsilon^{2})
\label{eq40}
\end{equation}

\noindent with,

\begin{equation}
P_{0}(\{q_{\mu}(0)\},\{p_{\mu}(0)\}) = \frac{1}{Z_{0}}\exp \Big(-\beta\sum_{\mu=1}^{N} \frac{p_{\mu}^{2}}{2m_{\mu}} + \sum_{\mu=1}^{N}\frac{1}{2} m_{\mu}\omega_{\mu}^{2} q_{\mu}^{2}\Big).
\label{eq41}
\end{equation}

\noindent The one-point average of the noise is calculated and, to order $\lambda \epsilon$, is found to deviate from zero by a constant. This is a feature of odd-powered nonlinearities. This feature also manifests itself in potentials where perturbative expansions can be carried out around some stable minimum that is not a point of symmetry of the potential landscape \cite{bhadra}. Simple inspection shows that only the noise kernel dependent on $\lambda \epsilon$ has terms \textit{even} in the initial values of the bath coordinates and contribute to the average of the noise. Thus, upto first order in $\epsilon$, $\langle \Gamma(t) \rangle$ becomes:

\begin{equation}
\langle{\Gamma(t)}\rangle = -\lambda\epsilon \sum_{\mu=1}^{N}\frac{1}{\beta m_{\mu}^{2} \omega_{\mu}^{4}}.
\label{eq42}
\end{equation}

\noindent Next, we turn to the two-time noise correlations, i.e., $\langle \Gamma(t)\Gamma(t{'} \rangle$. The noise kernel now has three distinct terms of different orders in $\lambda$ and $\epsilon$ and the non-vanishing terms to be evaluated upto second order in $\epsilon^{2}$ are

\begin{eqnarray}
\langle\Gamma(t)\Gamma(t') \rangle &=& \langle \Gamma _{\lambda}(t)\Gamma _{\lambda}(t')\rangle + \langle \Gamma _{\lambda \epsilon}(t)\Gamma _{\lambda \epsilon}(t')\rangle_{0} + \langle \Gamma _{\lambda^{2} \epsilon}(t)\Gamma _{\lambda^{2} \epsilon}(t')\rangle_{0}\nonumber\\
&&+ \langle \Gamma _{\lambda}(t)\Gamma _{\lambda \epsilon}(t')\rangle + \langle \Gamma _{\lambda}(t)\Gamma _{\lambda ^{2} \epsilon}(t')\rangle + \langle \Gamma _{\lambda \epsilon}(t)\Gamma _{\lambda^{2} \epsilon}(t')\rangle_{0} + (t \leftrightarrow t').
\label{eq43}
\end{eqnarray}

\noindent The reader is reminded of the distinction between the two averages in the above equation, $\langle \ldots \rangle_{0}$ and $\langle \ldots \rangle$, the latter containing the cubic nonlinearity upto order $\epsilon$ in the probability distribution. Rigorously calculated, the different correlations categorised according to the relevant order in perturbation are

\begin{eqnarray}
\langle \Gamma_{\lambda}(t)\Gamma_{\lambda}(t') \rangle &=& \frac{\lambda^{2}}{\beta}\sum_{\mu=1}^{N}\frac{c_{\mu}^{2}}{m_{\mu} \omega_{\mu}^{2}}\cos (\omega_{\mu}(t-t'))
\nonumber\\
&&= \frac{\lambda^{2}}{\beta}\gamma^{l}(t-t')
\label{eq44} \\
\langle \Gamma_{\lambda}(t)\Gamma_{\lambda^{2}\epsilon}(t') \rangle
&=& -\frac{2\lambda ^{3}\epsilon}{\beta}\int_{0}^{t'} ds \sum_{\mu=1}^{N}\frac{c_{\mu}^{3}}{m_{\mu}^{3}\omega_{\mu}^{4}}\sin(\omega_{\mu} (t'-s)\cos (\omega_{\mu} (t-s))G_{\mu}(s)
\nonumber\\
&&= \frac{\lambda ^{3}\epsilon}{\beta}\int_{0}^{t'} ds\Big[\sum_{\mu=1}^{N}\sin(\omega_{\mu} (s-t{'}))\gamma^{nl}_{\mu}(t-s)\Big],
\label{eq45}
\end{eqnarray}

\noindent where ${\gamma}^l$ and ${\gamma}^{nl}_{\mu}$ have been defined in Eqs.(\ref{eq32}) and (\ref{eq34}) respectively. Other correlation terms that reveal the signatures of nonlinearities explicitly are,

\begin{eqnarray}
\langle \Gamma_{\lambda}(t)\Gamma_{\lambda \epsilon}(t')\rangle
&=& \frac{\lambda^2 \epsilon^2}{\beta^2} \sum_{\mu=1}^N \frac{c_{\mu}^2}{6 m_{\mu}^4 {\omega}_{\mu}^8} \Big[ 4\{3 + 8\cos (\omega_{\mu}t') + 2\cos(2\omega_{\mu} t')\}.\cos(\omega_{\mu}t) \sin^2 \big( \frac{\omega_{\mu}t'}{2} \big)
\nonumber \\
&&+ \{3 \sin(\omega_{\mu}t') - 2 \sin(2\omega_{\mu} t') + \sin (3\omega_{\mu} t')\}. \sin(\omega_{\mu} t) \Big]
\label{eq46} \\
\langle \Gamma_{\lambda\epsilon}(t)\Gamma_{\lambda \epsilon}(t')\rangle_{0}
&=& \frac{\lambda^{2}\epsilon^{2}}{\beta^{2}} \sum_{\mu=1}^{N} \frac{1}{9m_{\mu}^{4} \omega_{\mu}^{8}} \sin \Big(\frac{\omega_{\mu} t'}{2}\Big)\sin \Big(\frac{\omega_{\mu} t}{2}\Big)\Big[36 \cos \Big(\frac{1}{2}\omega_{\mu} (t'-t)\Big)
\nonumber\\
&& + 9 \cos \Big(\frac{3}{2} \omega_{\mu} (t'-t)\Big)
\cos \left(\frac{5}{2} \omega_{\mu} (t'-t)\right) - 36 \cos \left(\frac{1}{2} \omega_{\mu} (t'+t)\right)
\nonumber\\
&& + 3 \cos \left(\frac{1}{2} \omega_{\mu} (5t' - 3t)\right)
+ 3 \cos \left(\frac{1}{2} \omega_{\mu} (3t' - 5t)\right)\Big]
\label{eq47} \\
\langle \Gamma_{\lambda \epsilon}(t)\Gamma_{\lambda^{2} \epsilon}(t')\rangle_{0}
&=& 0
\label{eq48} \\
\langle \Gamma_{\lambda^{2}\epsilon}(t)\Gamma_{\lambda^{2}\epsilon}(t')\rangle_{0}
&=& \frac{4\lambda^{4}\epsilon^{2}}{\beta}\int_{0}^{t} ds \int_{0}^{t'} du\Big[\sum_{\mu=1}^{N}\frac{c_{\mu}^{4}}{m_{\mu}^{5}\omega_{\mu}^{6}}\sin(\omega_{\mu} (t-s))
\nonumber\\
&&\times \sin(\omega_{\mu} (t'-u))\cos \omega_{\mu}(u-s) G_{\mu}(s)G_{\mu}(u)\Big].
\label{eq49}
\end{eqnarray}

\noindent These set of equations encompass the two-time noise-correlations in our problem. Here, we are compelled to take a critical look at the limits of validity of our strong coupling picture. These excess correlations due to nonlinear bath can serve as corrections to the usual equilibrium picture only if $\lambda\epsilon<1$ even when $\lambda \approx O(1)$ . These conditions still satisfy the strong system-bath coupling regime as long as the effect of $\epsilon$ is truly perturbative in nature. Hence, simple inspection reveals that the \textit{first equation [Eq.(\ref{eq44})] with right hand side quadratic in only $\lambda$ still remains the most dominant term of all}. However, if we were to consider $\lambda\epsilon>1$, the expansion of the GLE and the noise-correlations become divergent since it is a truly non-perturbative problem in both these parameters. That intuitively explains why an arbitrary strong system coupled to a generalised nonlinear bath probably does not fit into the picture of partitioning them into separate time-independent Hamiltonians, although some approaches have been discussed recently \cite{jarz,stras}; only a quadratic bath (both solvable and non-perturbative) is suitable in such a situation.\\

\noindent A closer look at Eqs.(\ref{eq44})-(\ref{eq49}) is in order. The first equation is the usual FDR of the second kind at zeroth order in $\epsilon$. It is quadratic in $\lambda$ on right hand side and proportional to absolute temperature $T$. The nonlinearity does not enter into this formula and is valid for any type of system-bath coupling with \textit{quadratic baths}. This is the equilibrium nature of the system-bath dynamics and as mentioned above remains the most dominant correlation. The second equation reveals a generalised FDR where the noise-correlation is proportional to a state-dependent damping kernel $\gamma_{mu}^{nl}$ (as defined before in Eq.(\ref{eq34})). Although, an extra time kernel is involved, time-translation invariance and linear temperature dependence is explicitly maintained. Through these two equations, the connection of damping and thermal noise in GLE (Eq.(\ref{eq30})) are complete and the orders of perturbation match exactly.\\

\noindent The next three equations (Eqs.(\ref{eq46})-(\ref{eq48})) express the effect of nonlinearities explicitly; these correlations therefore are symbolic of the self-interaction of the bath modes due to the presence of a nonlinear term and the system coupling to them. The hallmark of these terms is the dependence on higher powers of $\frac{1}{\beta}$ and $\frac{1}{\omega_\mu}$ along with an explicit breaking of time-translational invariance. The standard form of kernel $K(t-t{'})$ is violated in these correlations. However, the basic transformation $t \leftrightarrow t'$ holds good. The emergence of different time-scales in these equations are noteworthy. Finally, the last equation Eq.(\ref{eq49}) of the order $\lambda^{4}\epsilon^{2}$ represents a complex higher order friction kernel that can only be investigated on the left hand side of the GLE, Eq.(\ref{eq30}) once the second order corrections to bath-mode solutions are evaluated. We mention again that a small $\lambda$ approximation washes away all but Eq.(\ref{eq47}), which are correlations occurring due to a weak coupling between the system and a nonlinear bath.\\

\section{A comparison with other Caldeira-Leggett type models}

\noindent Here, we discuss the context of our work in light of the extensions of the Caldeira-Leggett model \cite{legg} that has been extensively studied and applied over the years. The GLE and the structure of generalised FDR are compared to those models and the subtle differences are investigated. As a starter, we revisit the Caldeira-Leggett Model with a quadratic bath and nonlinear state dependent coupling. The components of the full Hamiltonian are

\begin{eqnarray}
H_{S} &=& \frac{P^{2}}{2M}+V(X)
\nonumber\\
H_{B} &=& \sum_{\mu=1}^{N} \frac{p_{\mu}^{2}}{2m_{\mu}}+\sum_{\mu=1}^{N}\frac{1}{2} m_{\mu}\omega_{\mu}^{2} q_{\mu}^{2}
\nonumber\\
H_{SB} &=& -\lambda \sum_{\mu=1}^{N} c_{\mu}q_{\mu}(t)H(X(t)).
\label{eq50}
\end{eqnarray}

\noindent The GLE becomes

\begin{equation}
M\ddot{X} + \lambda^{2} H'(X(t))\int_{0}^{t} ds \gamma^{l}(t-s)H'(X(s))\dot{X}(s) + V'(X)
= H'(X(t)) \Gamma_{\lambda}
\label{eq51}
\end{equation}

\noindent where primes denote derivatives with respect to $X$. Two aspects are revealed in Eq.(\ref{eq51}): one, the origin of coloured noise in the right hand side, and two, a mixing of probe position and velocity in the dissipation function in the left hand side. The damping kernel and the noise terms are related through the standard equilibrium FDR,

\begin{equation}
\langle \Gamma_{\lambda}(t)\Gamma_{\lambda}(t') \rangle = \frac{\lambda^{2}}{\beta}\gamma^{l}(t-t').
\label{eq52}
\end{equation}

\noindent The important point to note here is that the  state-dependence of $H'(X(t))$  occurs twice in the product form of the $\lambda^{2}$ term on the left hand side of Eq.(\ref{eq49}). Thus neither a negative sign  nor any odd-powered nonlinearities occur in the GLE. \textit{Therefore, it can be concluded that such Caldeira-Leggett type Hamiltonians cannot describe any anti-damping, i.e., negative dissipation phenomena due to interaction with the environment}. Moreover, even with perturbative nonlinearities (which follow the same mathematical analysis as above) in the bath, the GLE does not reveal a nonlinear dissipation function like that in Eq.(\ref{eq30}) in the weak coupling regime \cite{cortes,hu}. The FDR in such instances remain intact in the zeroth order and picks up corrections only through the effect of weak bath nonlinearities. The class of noisy dynamical systems with antidamping or limit cycle characteristics cannot be constructed out of these models.\\

\noindent For completeness, we end by sketching similar issues when the coupling is nonlinear in bath coordinates as well, i.e.,

\begin{equation}
H_{SB} = -\lambda \sum_{\mu=1}^{N} c_{\mu}J(q_{\mu}(t))H(X(t)).
\label{eq53}
\end{equation}

\noindent The equation of motions for the system and the bath become,

\begin{eqnarray}
M\ddot{X} &=& -V'(X) + \sum_{\mu=1}^{N} \lambda H'(X(t))J(q_{\mu}(t))
\nonumber \\
m\ddot{q}_{\mu} &=& -m \omega^{2} q_{\mu} + \lambda X(t)\frac{\partial J(q_{\mu})}{\partial q_{\mu}}.
\label{eq54}
\end{eqnarray}

\noindent The equation of the bath mode cannot be solved with arbitrary coupling strength; the usual step would be to perform a perturbative expansion around the quadratic mode and then use the zeroth order solution to get the full inhomogeneous solution.

\begin{equation}
q(t) = q^{(0)} + \lambda q^{(1)} + O(\lambda ^{2}).
\label{eq55}
\end{equation}

\noindent \textit{Effectively, the system-bath coupling has to be treated perturbatively to make any progress}. Therefore, at the classical level even this model is handicapped by certain approximations which does not allow the non-perturbative treatment of strong system-bath coupling. Only through the subtle interplay of coupling and nonlinearity of the bath can new nonlinear dissipation functions and higher order correlations in the FDR be obtained as shown in the previous sections.

\section{Summary and Outlook}

\noindent In the present paper, we have worked on an extension of the Caldeira-Leggett open system model in a classical setting. The thermal bath has been considered to be perturbatively nonlinear; the system is,however,coupled arbitrarily strongly to the bath. Thus, the system-bath coupling can no longer be tackled via a perturbation expansion which leads to rich physics in the resulting GLE. Firstly, the position and velocity of the probe system gets intrinsically mixed in the dissipation function arising in the GLE. On one hand, the usual linear velocity-dependent damping kernel is obtained. Also, a second nonlinear damping function occurs that in the slow probe limit shows quadratic velocity dependence. Moreover, the sign of the system-bath coupling can play a crucial role and can be tuned to obtain anti-damping mechanism in the open system dynamics. In fact, a the noise force itself produces a piece that is nonlocal;it is a coarse-graining in time of a distribution of random kicks on the system. It has been shown explicitly that these features help us get a deeper picture than what is revealed by Caldeira-Leggett models studied throughout the literature.\\

\noindent Secondly, the FDR arising out of such a model exhibits interesting departures from standard equilibrium picture. At the zeroth order of perturbation in bath nonlinearity, the standard FDR is retained connecting linear local noise and linear dissipation function. In subsequent orders, the FDR takes up a generalised nonlocal form yet retaining the connection between the nonlinear damping and nonlinear noise. At this order a completely independent term arises due to the presence of nonlinearity of the bath; this we believe is the signature of the fluctuations due to the self-interactions of the bath modes. Finally at even higher orders, state dependent multiple time kernel correlations occur. Through this explicit calculation, the inherent non-equilibrium nature of the strongly coupled system dynamics is revealed.\\

\noindent As a future direction, we remark that the relatively new discipline of Quantum Thermodynamics has been a rich field of study for strong coupling dynamics, a sector in which many open questions remain and many debates and controversies continue to enrich our understanding. It will not be out of place to mention that the present state of understanding on this subject will be appropriately supplemented as well as strengthened by a visualization at the microscopic level, where evolution equations for dynamical variables along with their corresponding probabilistic pictures play a fundamentally important role. Recently, a new technique has been derived to handle stochastic thermodynamic problems in this regime and the approach presented in this paper bears a close resemblance to the generalised formalism laid out in \cite{stras}. In summary, this rigorous mathematical analysis with complicated realistic baths will hopefully shed light on quantum and classical aspects of non-equilibrium statistical mechanics, both for theorists and experimentalists.\\

\noindent\textbf{Acknowledgements:} The author is grateful to Professor Christian Maes (KU Leuven, Belgium) during his visit to Leuven in May 2017, for many motivating discussions and hospitality. The author is also grateful for useful discussions with Professor Abhishek Dhar (ICTS, India) during several meetings. His thesis advisor Dr.Dhruba Banerjee (Jadavpur University, India) is also sincerely acknowledged for careful reading and insights on the manuscript. Finally, he is thankful to UGC, Government of India, for financial support.

\section{Appendix}
\subsection{Explicit nonlinear terms for evaluation of perturbative expansion}

\noindent In the right hand side of Eq.(\ref{eq13a}) several constants had been introduced. Their structures are detailed below. The expression for ${[q^{(0)}(t)]}^2$ was

\begin{eqnarray}
{[q^{(0)}(t)]}^2 &=& A_{0} + A_{1} \cos (2\omega^{(0)}t) + A_{2} \sin (2\omega^{(0)}t)
\nonumber \\
&&+ \frac{2\lambda}{m\omega^{(0)}} F(t) \Big(q^{(0)} (0) \cos (\omega^{(0)}t) +  \frac{p^{(0)} (0)}{m \omega^{(0)}} \sin (\omega^{(0)}t)\Big) + \frac{\lambda^{2}}{m^{2}{{\omega^{(0)}}^2}} F^{2}(t).
\label{eq56}
\end{eqnarray}

\noindent Here

\begin{eqnarray}
A_{0} &=& -\frac{1}{2}\Big(q^{2}(0) +\frac{p^{2}(0)}{m^{2}{[\omega^{(0)}]}^2}\Big)
\label{eq56a} \\
A_{1} &=& -\frac{1}{2}\Big(q^{2}(0) -\frac{p^{2}(0)}{m^{2}{[\omega^{(0)}]}^2}\Big)
\label{eq56b} \\
A_{2} &=& -\frac{q(0)p(0)}{m\omega^{(0)}}.
\label{eq56c}
\end{eqnarray}

\noindent In Eq.(\ref{eq17}) the following constants appeared:

\begin{eqnarray}
B_{0} &=& \frac{1}{2}\Big(\frac{q^{2}(0)}{\omega^{(0)}} +\frac{p^{2}(0)}{m^{2}{[\omega^{(0)}]}^3}\Big)
\label{eq56d} \\
B_{1} &=& \frac{A_{2}}{2}
\label{eq56e} \\
B_{2} &=& \frac{A_{1}}{2\omega_{0}}.
\label{eq56f}
\end{eqnarray}

\subsection{Transposing one-particle bath to a realistic many-particle scenario}

\subsubsection{Nonlinear noise-kernel}

\noindent The noise-kernels for many-particle bath, appearing on the right hand side of Eq.(\ref{eq30}), have the following structures:

\begin{eqnarray}
\Gamma_{\lambda}(t)&=& \lambda \sum_{\mu=1}^{N}c_{\mu}\Big(q_{\mu}(0) \cos \omega_{\mu} t + \frac{p_{\mu}(0)}{m_{\mu} \omega_{\mu}} \sin \omega_{\mu} t\Big)
\label{eq57a} \\
\Gamma _{\lambda \epsilon}(t)&=&\sum_{\mu=1}^{N}\frac{\lambda \epsilon c_{\mu}}{m_{\mu}\omega_{\mu}} \Big[B_{\mu 0}\Big(\cos \omega_{\mu} t - 1 \Big) + B_{\mu 1}\Big(\sin3\omega_{\mu} t +\frac{2}{3}\sin 2\omega_{\mu} t -\sin\omega_{\mu} t\Big)\nonumber\\
&& + B_{\mu 2}\Big(\cos3\omega_{\mu} t +\frac{2}{3}\cos2\omega_{\mu} t -\cos\omega_{\mu} t\Big)\Big]
\label{eq57b} \\
\Gamma _{\lambda^{2} \epsilon}(t)&=& -\sum_{\mu=1}^{N}\frac{2\lambda^{2} \epsilon c_{\mu}^2}{m_{\mu}^{2}\omega_{\mu}^{2}}\Big[\int_{0}^{t} dt'\sin (\omega _{\mu}(t-t')) G_{\mu}(t')\Big( q_{\mu}(0) \cos \omega_{\mu} t' + \frac{p_{\mu}(0)}{m_{\mu} \omega_{\mu}}\sin \omega_{\mu} t'\Big)\Big]
\label{eq57c}
\end{eqnarray}

\subsubsection{Partition function and Probability distribution with nonlinearity}

\noindent The method of ensemble averaging described between Eqs.(\ref{eq35}) and (\ref{eq40}) is summarized below.

\begin{eqnarray}
Z_{0} &=& \int d\Omega \Big[\exp \Big(-\beta\sum_{\mu=1}^{N} \frac{p_{\mu}^{2}}{2m_{\mu}}+\sum_{\mu=1}^{N}\frac{m_{\mu}\omega_{\mu}^{2}}{2} q_{\mu}^{2}\Big)\Big]\nonumber\\
&=& \prod_{i=1}^{N}{\frac{2\pi}{\beta\omega_{\mu}}}
\label{eq58} \\
Z' &=& \int \exp \Big(-\beta H_{B}\Big)d\Omega\nonumber\\
&=& \int \exp \Big(-\beta(H_{0}+\epsilon H_{n})\Big)d\Omega\nonumber\\
&\approx& \int \Big(1-\beta \epsilon  H_{n}\Big)\exp \Big(-\beta H_{0}\Big)d\Omega\nonumber\\
&=& Z_{0}\Big(1-\beta \epsilon \langle{H_{n}}\rangle_{0}\Big)
\label{eq59} \\
P'(\{q_{\mu}(0)\},\{p_{\mu}(0)\}) &=& \frac{1}{Z'}\Big(-\beta(H_{0}+\epsilon H_{n})\Big)\nonumber\\
&=& \frac{1}{Z_{0}}\Big(1-\beta \epsilon  H_{n}\Big)\exp \Big(-\beta H_{0}\Big)+O(\epsilon^{2})\nonumber\\
&\approx& (1-\beta\epsilon H_{n})P_{0}(\{q_{\mu}(0)\},\{p_{\mu}(0)\}).
\label{eq60}
\end{eqnarray}

\subsubsection{The one-point average to order $\epsilon$}

\noindent The right hand side of Eq.(\ref{eq42}) is arrived at in the following way:

\begin{eqnarray}
\langle{\Gamma(t)}\rangle &=& \langle{\Gamma_{\lambda\epsilon}(t)}\rangle_{0} - \beta\epsilon\langle{H_{n}\Gamma_\lambda(t)}\rangle_{0}\nonumber\\
&=& \Big[\lambda\epsilon \sum_{\mu=1}^{N}\frac{1}{\beta m_{\mu}^{2} \omega_{\mu}^{4}}\Big(\cos(\omega_{\mu} t)-1\Big)\Big] - \Big[\lambda\epsilon \sum_{\mu=1}^{N}\frac{1}{\beta m_{\mu}^{2} \omega_{\mu}^{4}}\Big(\cos(\omega_{\mu} t)\Big]\nonumber\\
&=& -\lambda\epsilon \sum_{\mu=1}^{N}\frac{1}{\beta m_{\mu}^{2} \omega_{\mu}^{4}}.
\label{eq61}
\end{eqnarray}

\noindent All other terms at this order vanish due to even-odd pairing of the initial bath coordinates.

\subsubsection{Two-time Noise-correlations: some sample integrals}

\noindent The integrals involved in the evaluation of the two-time correlation functions between Eqs.(\ref{eq44}) and (\ref{eq49}) are cumbersome but straightforward Gaussian integrations. Two sample integrals are elaborated below. The integral in Eq.(\ref{eq44}) is evaluated as,

\begin{eqnarray}
\langle \Gamma _{\lambda}(t)\Gamma _{\lambda}(t{'}) \rangle &=& \frac{\lambda^{2}}{Z_{h}} \int d\Omega \exp \Big[ -\beta \Big(\sum_{\sigma=1}^{N} \frac{p_{\sigma}^{2}}{2m_{\sigma}} + \frac{m_{\sigma}\omega_{\sigma}^{2}}{2} q_{\sigma}^{2}\Big)\Big]
\nonumber \\
&&\times \sum_{\mu,\nu=1}^{N}\Big[c_{\mu}\Big(q_{\mu}(0) \cos \omega_{\mu} t + \frac{p_{\mu}(0)}{m_{\mu} \omega_{\mu}} \sin \omega_{\mu} t\Big)
c_{\nu}\Big(q_{\nu}(0) \cos \omega_{\nu} t' + \frac{p_{\nu}(0)}{m_{\nu} \omega_{\nu}} \sin \omega_{\nu} t'\Big)\Big] \nonumber\\
&=& \frac{\lambda^{2}}{\beta}\sum_{\mu=1}^{N}\frac{c_{\mu}^{2}}{m_{\mu} \omega_{\mu}^{2}}\cos (\omega_{\mu}(t-t')),
\label{eq62}
\end{eqnarray}

\noindent where the different subscripts get absorbed into one due to orthogonality relations of the trigonometric functions. The integral in Eq.(\ref{eq45}) is evaluated in a similar way as follows:

\begin{eqnarray}
\langle \Gamma_{\lambda}(t)\Gamma_{\lambda^{2}\epsilon}(t') \rangle
&=& -\frac{2\lambda^{3} \epsilon}{Z_{h}}\int d\Omega \exp \Big[ -\beta \Big(\sum_{\sigma=1}^{N} \frac{p_{\sigma}^{2}}{2m_{\sigma}} + \frac{m_{\sigma}\omega_{\sigma}^{2}}{2} q_{\sigma}^{2}\Big)\Big]
\nonumber \\
&&\times \sum_{\mu,\nu=1}^{N}\Big[c_{\mu}\Big(q_{\mu}(0) \cos \omega_{\mu} t + \frac{p_{\mu}(0)}{m_{\mu} \omega_{\mu}} \sin \omega_{\mu} t\Big)
\nonumber\\
&&\times \frac{c_{\nu}^{2}}{m_{\nu} \omega_{\nu}}\Big(\int_{0}^{t'} ds\sin (\omega_{\nu}(t'-s)) G_{\nu}(s)\Big( q_{\nu}(0) \cos \omega_{\nu} s + \frac{p_{\nu}(0)}{m_{\nu} \omega_{\nu}}\sin \omega_{\nu} s\Big)\Big]
\nonumber\\
&=& -\frac{2\lambda^{3} \epsilon}{Z_{h}}\int_{0}^{t'} ds\sum_{\mu,\nu=1}^{N}\Big[\sin (\omega_{\nu}(t{'}-s)) G_{\nu}(s)
\nonumber \\
&&\times \int d\Omega\exp \Big\{ -\beta \Big( \sum_{\mu=1}^{N} \frac{p_{\mu}^{2}}{2m_{\mu}} + \frac{m_{\mu}\omega_{\mu}^{2}}{2} q_{\mu}^{2}\Big)\Big\}
\nonumber\\
&&\times \Big\{c_{\mu}\Big(q_{\mu}(0) \cos \omega_{\mu} t + \frac{p_{\mu}(0)}{m_{\mu} \omega_{\mu}} \sin \omega_{\mu} t\Big).\frac{c_{\nu}^{2}}{m_{\nu} \omega_{\nu}} \Big( q_{\nu}(0) \cos \omega_{\nu} s + \frac{p_{\nu}(0)}{m_{\nu} \omega_{\nu}}\sin \omega_{\nu} s\Big)\Big\}\Big]
\nonumber\\
&=& -\frac{\lambda^{3} \epsilon}{\beta}\int_{0}^{t'} ds\sum_{\mu=1}^{N}\Big[\sin (\omega_{\mu}(t'-s)) G_{\mu}(s)\frac{2c_{\mu}^{3}}{m_{\mu}^{3} \omega_{\mu}^{4}}\cos (\omega_{\mu}(t-s))\Big]
\nonumber\\
&=& \frac{\lambda^{3}\epsilon}{\beta}\int_{0}^{t'} ds\Big[\sum_{\mu=1}^{N}\sin(\omega_{\mu} (s-t'))\gamma^{nl}_{\mu}(t-s)\Big]
\label{eq63}
\end{eqnarray}

\noindent where, again, the indices $\sigma$ and $\nu$ get absorbed into $\mu$ due to orthogonality relations of the trigonometric functions.

\end{document}